\documentclass[twocolumn,prb]{revtex4-2}
\usepackage[T1]{fontenc}
\usepackage[utf8]{inputenc}
\usepackage{float}
\usepackage{graphicx}
\usepackage{xcolor}
\usepackage{hyperref}
\usepackage{babel}
\usepackage{physics}
\usepackage{amsmath,amssymb,amsfonts}
\usepackage{appendix}
\usepackage{dsfont} 
\usepackage{blindtext}
\usepackage{tikz}

\definecolor{plotred}{HTML}{FF2C00}
\definecolor{plotorange}{HTML}{FF9500}
\definecolor{plotgreen}{HTML}{00B945}

\begin{document}

\title{\vspace{-2cm}\bfseries Third-order topological insulator induced by disorder}
\author{Hugo Lóio$^{1,2}$, Miguel Gonçalves$^1$, Pedro Ribeiro$^{1,3}$, Eduardo V. Castro$^{1,3,4}$}
\affiliation{$^{1}$CeFEMA, LaPMET, Instituto Superior Técnico, Universidade de Lisboa,
Av. Rovisco Pais, 1049-001 Lisboa, Portugal}
\affiliation{$^{2}$Laboratoire de Physique Th\'eorique et Mod\'elisation, CNRS UMR 8089, CY Cergy Paris Universit\'e, 95302 Cergy-Pontoise Cedex, France}
\affiliation{$^{3}$Beijing Computational Science Research Center, Beijing 100084,
China}
\affiliation{$^{4}$Centro de F\'{\i}sica das Universidades do Minho e Porto, LaPMET, Departamento
de F\'{\i}sica e Astronomia, Faculdade de Ciências, Universidade do
Porto, 4169-007 Porto, Portugal}

\begin{abstract}
  We have found the first instance of a third-order topological Anderson insulator (TOTAI). This disorder-induced topological phase is gapped and characterized by a quantized octupole moment and topologically protected corner states, as revealed by a detailed numerically exact analysis. We also find that the disorder-induced transition into the TOTAI phase can be analytically captured with remarkable accuracy using the self-consistent Born approximation. For a larger disorder strength, the TOTAI undergoes a transition to a trivial diffusive metal, that in turn becomes an Anderson insulator at even larger disorder. 
  Our findings show that disorder can induce third-order topological phases in 3D, therefore extending the class of known higher-order topological Anderson insulators.
\end{abstract}

\maketitle

\section{Introduction}

In symmetry-protected topological (SPT) phases of matter, such as topological insulators (TIs), non-trivial bulk topology leads to protected gapless excitations on the system's boundary \cite{hasan2010colloquium, qi2011topological,moore2010birth,shen2012topological}.
These edge-states have exotic, disorder-robust properties with promising applications for quantum computation \cite{nayak2008non,freedman2003topological,stern2013topological}.
SPT phases of matter are classified in the \textit{ten fold way} \cite{schnyder2008classification}, based on the discrete symmetries (time-reversal, charge-conjugation and chiral) that constrain the system's Hamiltonian.
Spatial symmetries of crystalline nature may also be encountered, producing topological crystalline insulators (TCIs) \cite{fu2011topological,neupert2018topological}.
Recently, TIs have been generalized to higher-order topological insulators (HOTIs), where the bulk-boundary correspondence applies to the $(d-n)$ dimensional boundary, for a $d$-dimensional, $n$th-order topological insulator \cite{schindler2018higher,khalaf2018higher,park2019higher,wang2021higher,franca2018anomalous,chen2020higher}. 
HOTIs were first demonstrated in the Benalcazar-Bernevig-Hughes (BBH) models \cite{benalcazar2017quantized, benalcazar2017electric}, where the topological invariant corresponds to quantized bulk quadrupole or octupole electric moments respectively in a 2D second-order topological insulator (SOTI) and 3D third-order topological insulator (TOTI), with protected corner states.
In the BBH models, the topological properties are protected by spatial symmetries, rendering them an extension of the TCIs.

Many experimental implementations of HOTIs have since been found, first in classical metamaterial analogues like mechanical metamaterials \cite{serra2018observation}, electric circuits \cite{imhof2018topolectrical,PhysRevB.100.201406}, coupled microwave resonators \cite{peterson2018quantized}, photonic waveguides \cite{Dutt2020}; and later even in solid-state materials \cite{Schindler2018,shumiya2022evidence,lee2023spinful}.
In any practical realization of a system, disorder is present, e.g. due to defects in manufacturing and can even be tuned in metamaterials.
Disorder has a profound impact on quantum transport due to Anderson localization of electronic wave functions \cite{anderson1958absence,lagendijk2009fifty}. This gives rise to Anderson insulators, that can have gapless excitations in contrast with conventional (gapped) band insulators \cite{evers2008anderson}.
It is generally known that TIs are robust against symmetry-preserving disorder.
Still, with enough disorder, it is possible to suppress topological phases. Remarkably, increasing disorder can also induce topological transitions from trivial to topological phases, giving rise to Topological Anderson insulators (TAIs)~\cite{li2009topological,groth2009theory}, which have been experimentlly realized recently in different platforms ~\cite{meier2018observation,Stutzer2018,Liu2020}.

The  concept of TAIs was recently extended  to higher-order topological Anderson insulators (HOTAIs) in Ref.~\cite{Li2020}, where a 2D SOTI was obtained by adding chiral-symmetric disorder to the 2D-BBH model. This result establishes  chiral symmetry as a sufficient symmetry to protect the HOTAI phases, even when the crystalline symmetries are broken by disorder. 
A full phase diagram was obtained in Ref.~\cite{yang2021higher} for a system that can be mapped to the 2D-BBH model. It was found that the disorder-induced SOTI comes in two varieties with increasing disorder: the gapped and gapless HOTAI phases, followed by a Griffiths phase.
Noteworthy, the classical analogue of a 2D HOTAI was recently experimentally observed using electric circuits~\cite{PhysRevLett.126.146802}, where disorder can be tuned.
A disorder-driven 3D SOTI was also found in amorphous systems with structural disorder~\cite{Agarwala2020, wang2021structural, Peng2022}.

In this work, we  find the first instance of a disorder-induced third-order topological Anderson insulator (TOTAI). 
Our conclusions are drawn from from the numerical analysis of the interplay between topology and chiral symmetry preserving disorder in the 3D-BBH model. 
The TOTAI phase is gapped and undergoes a transition into a trivial (gapless) diffusive metal (DM) with increasing disorder. At significantly larger disorder, it turns into an Anderson insulator (AI).
The gapless HOTAI phase and the Griffiths phase are absent, in contrast to the disordered 2D-BBH model \cite{yang2021higher}.  The detailed topological, spectral, and localization properties of the different phases found are summarized in Fig.$\,$\ref{fig_phases_scheme} and Tab.$\,$\ref{tab_phases}, and will be justified in detail below.

This paper is structured as follows.
In Sec. \ref{sec_methods}, we present the model and the topological invariants that characterize non-trivial phases, and which we compute numerically.
Detailed numerical results are presented in Sec. \ref{sec_results}, which allowed for the full description of the phase diagram of the model.
We also analytically capture the disorder-induced topological phase transition using the self-consistent Born approximation.
In Sec. \ref{sec_conclusions} we discuss our results and their implications.



\section{Model and methods}\label{sec_methods}

\begin{figure}
  \resizebox{\columnwidth}{!}{
    \begin{tikzpicture}
      \node[draw=none,fill=none] at (0,0){\includegraphics[width=0.6\columnwidth]{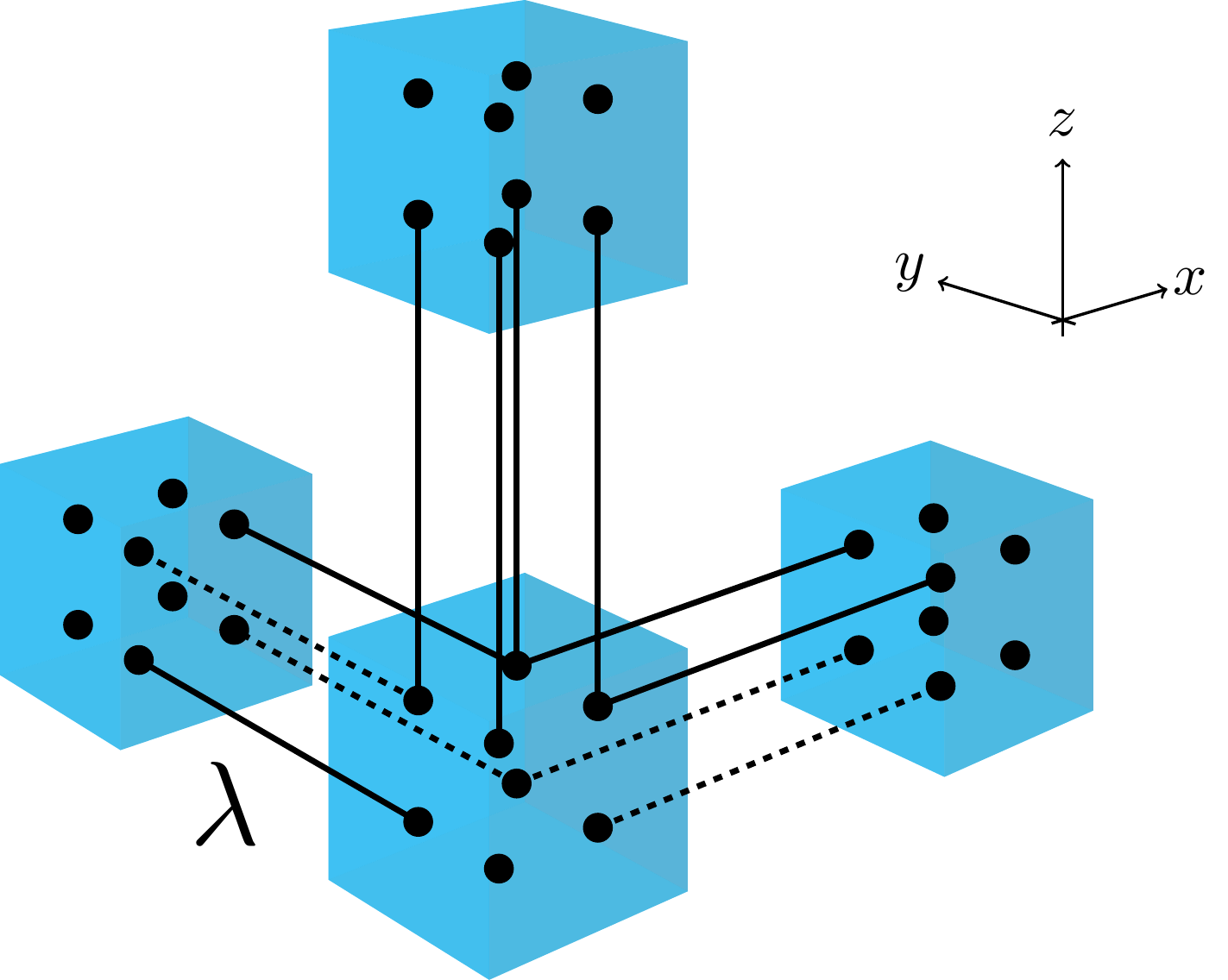}};
      \node at (-0.25\columnwidth,1.6) {\textbf{(a)}};
      \node[draw=none,fill=none] at (0.5\columnwidth,-0.2){\includegraphics[width=0.4\columnwidth]{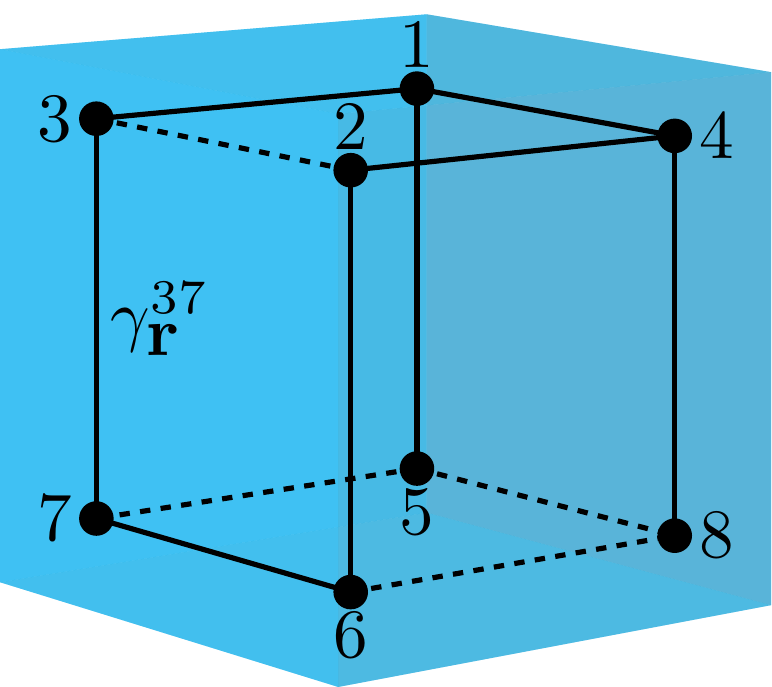}};
      \node at (0.35\columnwidth,1.6) {\textbf{(b)}};
      \node[draw=none,fill=none] at (0.2\columnwidth,-3.7){\includegraphics[width=0.4\columnwidth]{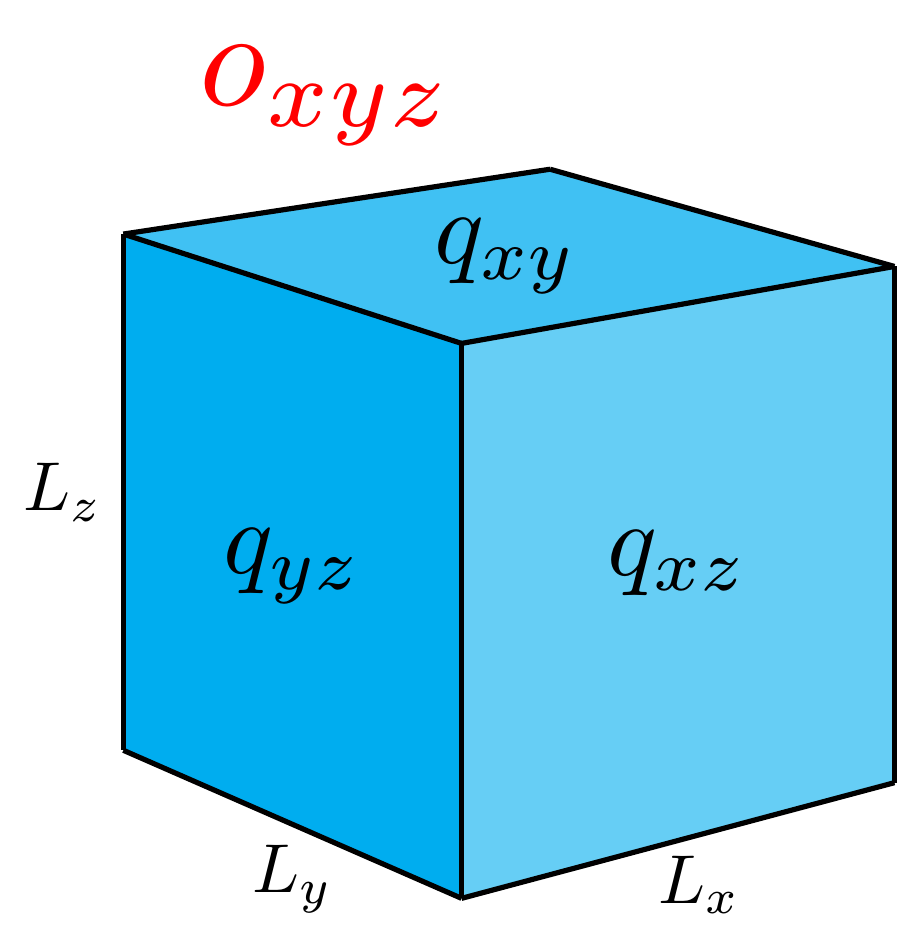}};
      \node at (0,-2.5) {\textbf{(c)}};
    \end{tikzpicture}
  }
  \caption{
    (a,b) Schematics of the 3D-BBH model with disorder. 
    In (a) only the inter-cell hoppings are shown, whilst in (b) the intra-cell hoppings are presented.
    Dotted lines correspond to negative signs in the clean hopping amplitudes.
    (c) Schematics of full system with bulk octupole moment $o_{xyz}$, boundary quadrupole moments $q_{ij}$ and size $L_i$ in directions $i,j \in \{x,y,z\}$.
  }
  \label{fig_scheme}
\end{figure}
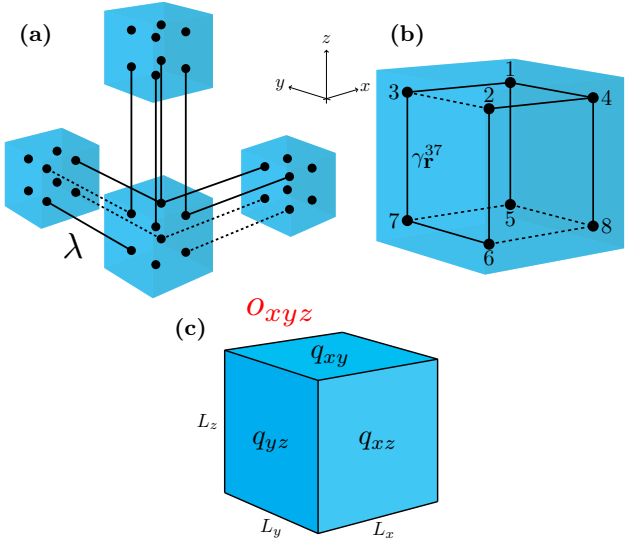

\textit{Model.---} The model under consideration is the 3D-BBH model \cite{benalcazar2017quantized}, generalized with disorder in the intra-cell hopping amplitudes, as illustrated in Fig. \ref{fig_scheme}(a,b).
The corresponding tight-binding Hamiltonian is given by
\begin{equation}\label{eq_ham}
  \hat{H} = \sum_{\textbf{r}} \left[ \hat{c}_{\textbf{r}}^\dagger \Gamma_{\textbf{r}} \hat{c}_{\textbf{r}} + \sum_{i \in \{x,y,z\}} \left( \hat{c}^\dagger_{\textbf{r}} \Lambda_i \hat{c}_{\textbf{r} + \textbf{e}_i} + H.c. \right) \right] \, , 
\end{equation}
where $\hat{c}^\dagger_{\textbf{r}} = (\hat{c}_{\textbf{r}1}^\dagger \ \hat{c}_{\textbf{r}2}^\dagger \ \dots \ \hat{c}_{\textbf{r}8}^\dagger) $, $\hat{c}_{\textbf{r}\alpha}^\dagger$ creates a particle at the $\alpha$-th site of cell $\textbf{r}$, and the hopping matrices are given by
\begin{equation}\label{eq_hop_mat}
  \begin{split}
    [\Gamma_\textbf{r}]_{ij} & =  \gamma_\textbf{r}^{ij}[\sigma_z \otimes (\sigma_x \otimes \mathds{1} - \sigma_y^{\otimes 2}) + \sigma_x \otimes \mathds{1}^{\otimes 2}]_{ij} \, , \\
    \Lambda_x & = \frac{\lambda}{2}\mathds{1} \otimes (\sigma_x \otimes \mathds{1} +i\sigma_y \otimes \sigma_z)  \, , \\
    \Lambda_y & = \frac{\lambda}{2}\mathds{1} \otimes i\sigma_y \otimes (\sigma_x  +i\sigma_y )  \, , \\
    \Lambda_z & = \frac{\lambda}{2}(\sigma_x + i\sigma_y) \otimes \mathds{1}^{\otimes 2} \, ,
  \end{split}
\end{equation}
where $\{ \mathds{1}, \sigma_x, \sigma_y, \sigma_z \}$ is the set of $2\times2$ identity and Pauli matrices. We set $\lambda = 1$ so that the energy is measured in units of $\lambda$. 
The intra-cell hopping amplitudes are (up to a sign as indicated in Fig. \ref{fig_scheme}(b) and Eq.~\eqref{eq_hop_mat}) $\gamma_\textbf{r}^{ij} = \gamma + W \Delta^{ij}_\textbf{r}$, where $W$ is the disorder strength and $\Delta_\textbf{r}^{ij} = \Delta_\textbf{r}^{ji}$ are uniformly  distributed random variables in the interval $[-\frac{1}{2}, \frac{1}{2}]$ without correlation. In our finite-size calculations, we consider cubic systems of size $L_x = L_y = L_z = L$.

In the clean limit, $W=0$, we have  $H\rightarrow H_0$, $\Gamma_\textbf{r} \rightarrow \Gamma_0$ and translational invariance allows us to express the Hamiltonian in reciprocal space as,
\begin{equation}
  \resizebox{0.90\columnwidth}{!}{
    $
    \begin{split}
      H_0(\textbf{k}) = &\, \sigma_z \otimes \left[\sigma_x \otimes \mathds{1} (\cos(k_x) + \gamma) - \sigma_y \otimes \sigma_z \sin(k_x)\right] \\
      - &\, \sigma_z \otimes \sigma_y \otimes \left[ \sigma_y (\cos(k_y) + \gamma) + \sigma_x \sin(k_y)\right] \\
      + & \left[ \sigma_x (\cos(k_z) + \gamma) - \sigma_y \sin(k_z)\right] \otimes \mathds{1}^{\otimes 2} \\
    \end{split}
    \, .
    $
  }
  \label{eq:clean_H}
\end{equation}
The topological properties depend on the value of the parameter $\gamma$, as discussed next.

\textit{Topological properties.---}
When $|\gamma| < 1$ ($|\gamma|  > 1$), the clean Hamiltonian in Eq.$\,$\eqref{eq:clean_H} is in a topological (trivial) phase with quantized octupole moment $o_{xyz}$. 
In reciprocal space, $o_{xyz}$ may be computed by the \textit{nested Wilson loop} method, where the spatial reflection and inversion symmetries of the clean system, along with time-reversal, charge-conjugation, and chiral symmetries were shown to protect the topology \cite{benalcazar2017quantized}. 
In real space, $o_{xyz}$ is computed through many-body electric multipole operators \cite{wheeler2019many,kang2019many,resta1998quantum}.
Since this involves finding the ground state of the system, it is computationally demanding to do it in 3D. 
However, in the topological phase, we also expect to find quantized quadrupole moments $q_{xy}, q_{xz}, q_{yz}$ in the 2D-boundaries of the insulator, as illustrated in  Fig. \ref{fig_scheme}(c), allowing for the definition of the topological invariant
\begin{equation}
  Q = 8 \left| q_{xy} q_{xz} q_{yz} \right| \ , 
  \label{Q_invariant}
\end{equation}
where each quadrupole moment is expressed as
\begin{equation}
  q_{ab} = \left[ \frac{1}{2 \pi} \textnormal{Im} \log \bra{\Psi_{c}} \mathcal{U}_{ab} \ket{\Psi_{c}}  - q^{(0)}_{ab} \right] \textnormal{mod } 1 \ ,
\end{equation}
with
\begin{equation}
  \mathcal{U}_{ab} = \exp \left(\frac{2\pi i \sum_{j=1}^{N_{\textnormal{occ}}} \hat{r}^j_a \hat{r}^j_b}{L_a L_b} \right) \ ,
\end{equation}
for $c \neq a \neq b$, where $\hat{r}_a^j$ is the position operator in direction $a=x,y,z$ for electron $j$ and $N_\textnormal{occ} = 2L_aL_b$ the number of occupied states in the boundary $ab$, with $L_a$ the number of unit cells in direction $a$.
$q^{(0)}_{ab} = \frac{1}{2}\sum^{N_\textnormal{a}}_{j=1} r_a^j r_b^j/(L_a L_b)$ is the contribution from the positive background charge, taking into account that the sample is electrically neutral with $N_\textnormal{a} = 2N_{\textnormal{occ}}$ atomic orbitals in the boundary.
$\ket{\Psi_c}$ is the boundary many-body ground state obtained from the effective Hamiltonian $H_c = -{G_{N_c}^{c}}(E = 0)^{-1}$, with $N_c = 2L_c$.
$G_{N_c}^{c}$ is the boundary Green's function \cite{peng2017boundary} that can be computed by dividing the Hamiltonian matrix into 2D layers in the direction $c$ and solving the following Dyson equation,
\begin{equation}
  G_{n}^c = (E - h_{n}^c - V_{n-1}^c G_{n-1}^c {V_{n-1}^{c \, \dagger}}) \ ,
\end{equation}
where $h_{n}^c$ is the Hamiltonian of the $n$th-layer, and the $(n-1)$th-layer couples to the $n$th-layer through matrix $V_{n-1}^c$.
The reduced Hilbert space dimensionality of each layer allows for reaching far larger system sizes when computing $Q$ than by computing the bulk octupole moment through many-body electric multipole operators.

In the disordered system, spatial crystalline symmetries are broken.
However, the system is still chiral symmetric, since it is decomposable in sub-latices that do not possess any hopping terms between themselves.
We will see that this symmetry suffices to protect the topology.
Chiral symmetry is also preserved in the effective boundary Hamiltonian.
The quadrupole moments are known to be quantized by chiral symmetry \cite{yang2021higher}, which means that, in each realization of disorder, $Q$ is quantized to $0$ or $1$.

\textit{Spectral properties.---} 
To study the spectral properties of the different phases, we computed the energy gap using exact diagonalization and the density of states (DOS), $\rho(E) = \frac{1}{D} \sum_{k=0}^{D-1}\delta(E-E_k)$, where $D$ is the Hilbert space dimension and ${E_k}$ are the single-particle eigenenergies. 
For an efficient calculation of the DOS, we employed the kernel polynomial method (KPM) \cite{weisse2006kernel}.
In all our KPM calculations, we evaluated the trace stochastically over a single random state and used the Jackson kernel.
A related quantity that can also be computed with the KPM is the local density of states (LDOS), $\rho(E, \textbf{r}) = \sum_{k=0}^D \sum_\alpha \left| \psi_k(\textbf{r},\alpha) \right|^2 \delta(E-E_k)$, where $\psi_k(\textbf{r},\alpha)$ is the $k$th eigenfunction evaluated at unit cell $\textbf{r}$ and orbital $\alpha$. 
We used this quantity to inspect the existence of localized corner states, to complement the analysis on the topological properties.

\textit{Localization properties.---}
Finally, we also study the localization properties of the eigenstates by evaluating their localization length, the average level-spacing ratio (LSR), the inverse participation ratio (IPR) and the fractal dimension.

The normalized localization length $\Lambda = \lambda/L$, where $\lambda$ is the localization length along the $z$ direction and $L = L_x = L_y$, was computed using the transfer matrix method (TMM) \cite{hoffmann2002computational,kramer1996transfer}. For extended states, $\Lambda$ increases with $L$, while for localized states, $\Lambda \rightarrow 0$, since $\lambda$ is finite. At critical points, $\lambda \sim L$ and therefore $\Lambda \sim L^0$.

The LSR is given by
\begin{equation}
  \textnormal{LSR} = \frac{1}{n-2} \sum_{i=1}^{n-2} \frac{\min(\delta_i, \delta_{i+1})}{\max(\delta_i, \delta_{i+1})} \ ,
\end{equation}
where $\delta_i = E_{i+1} - E_{i}$ are the spacings between $n$ eigenenergies $E_i$ sorted in ascending order.
The energy gap spacing is not included.
We expect the energy level spacings of localized eigenstates to follow Poisson statistics, in which case $\textnormal{LSR} \approx 0.386$. 
For diffusive extended states, the level spacings follow the Gaussian Orthonormal Ensemble (GOE) probability distribution, corresponding to $\textnormal{LSR} \approx 0.530$ \cite{oganesyan2007localization}.

The IPR \cite{wegner1980inverse} is expressed as
\begin{equation}
  \textnormal{IPR} = \frac{1}{n} \sum_{i=1}^n \sum_\textbf{r} \left(\sum_\alpha \left| \psi_i(\textbf{r}, \alpha) \right|^2 \right)^2 \ , 
\end{equation}
where $\psi_i(\textbf{r},\alpha)$ is the amplitude of the $i$th eigenfunction at unit cell $\textbf{r}$ and orbital $\alpha$. The IPR scales with system size as $\textnormal{IPR} \propto L^{-D_2}$, where $D_2$ is the (real-space) fractal dimension given by $D_2=3$ for extended states, $D_2=0$ for localized states and $0<D_2<3$ for fractal or multifractal states \cite{evers2008anderson}.

\section{Results}\label{sec_results}

\begin{figure}
  \centering
  \includegraphics[width=3in]{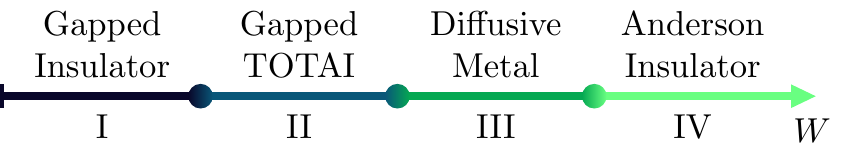}
  \caption{ 
  Schematic phase diagram as a function of $W$ for $\gamma = 1.1$.
  }
  \label{fig_phases_scheme}
\end{figure}

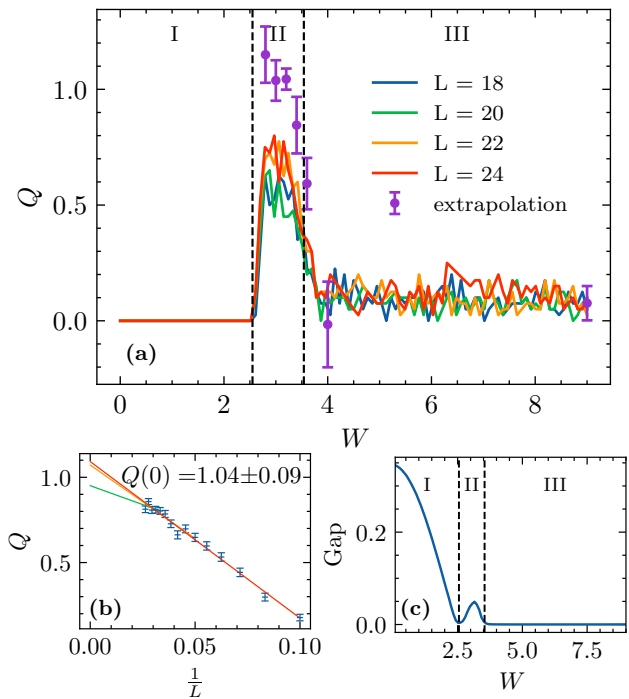
\begin{figure}
  \centering
  \resizebox{\linewidth}{!}{
    \begin{tikzpicture}
      \node[draw = none, fill = none] at (-0.01\columnwidth,0){\includegraphics[width=\columnwidth]{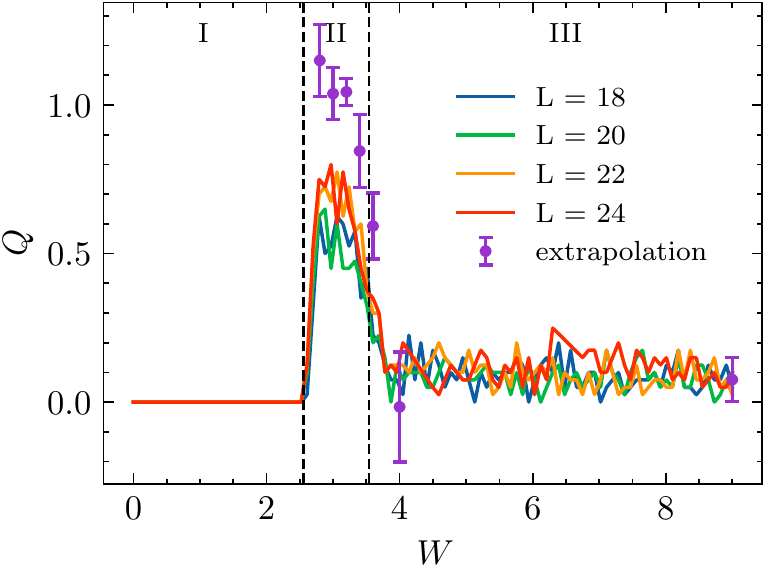}};
      \node at (-0.3\columnwidth,-1.8) {\textbf{(a)}};
      \node[draw = none, fill = none] at (-0.26\columnwidth,-5){\includegraphics{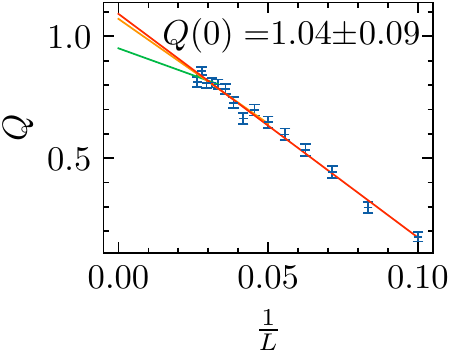}};
      \node at (-0.36\columnwidth,-5.5) {\textbf{(b)}};
      \node[draw = none, fill = none] at (0.26\columnwidth,-5){\includegraphics{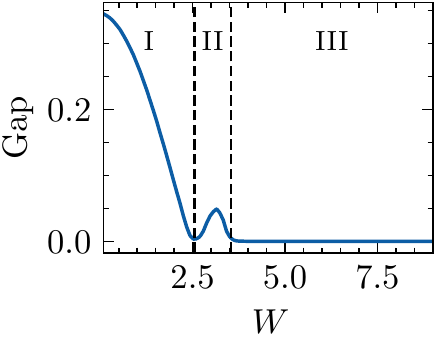}};
      \node at (0.16\columnwidth,-5.5) {\textbf{(c)}};
    \end{tikzpicture}
  }
  \caption{
    (a) Topological phase diagram obtained from the topological invariant $Q$ defined in Eq.$\,$\eqref{Q_invariant} with respect to the disorder strength $W$. 
    For the lines with fixed size, $Q$ was averaged with 40 disorder realizations. 
    To compute the extrapolated points at some selected values of $W$, $Q$ was averaged over 400 disorder realizations.
    In (b), an example of the extrapolation is shown for $W = 3$.
    (c) Bulk energy gap computed from exact diagonalization for a system size $L=20$ and averaging over $200$ disorder realizations, as a function of $W$.
  }
  \label{fig_phase_diag}
\end{figure}

\begin{figure}
  \centering
  \resizebox{\columnwidth}{!}{
    \def\width{1.7in}
    \begin{tikzpicture}
      \node[draw = none, fill = none] at (0,0){\includegraphics[width=\width]{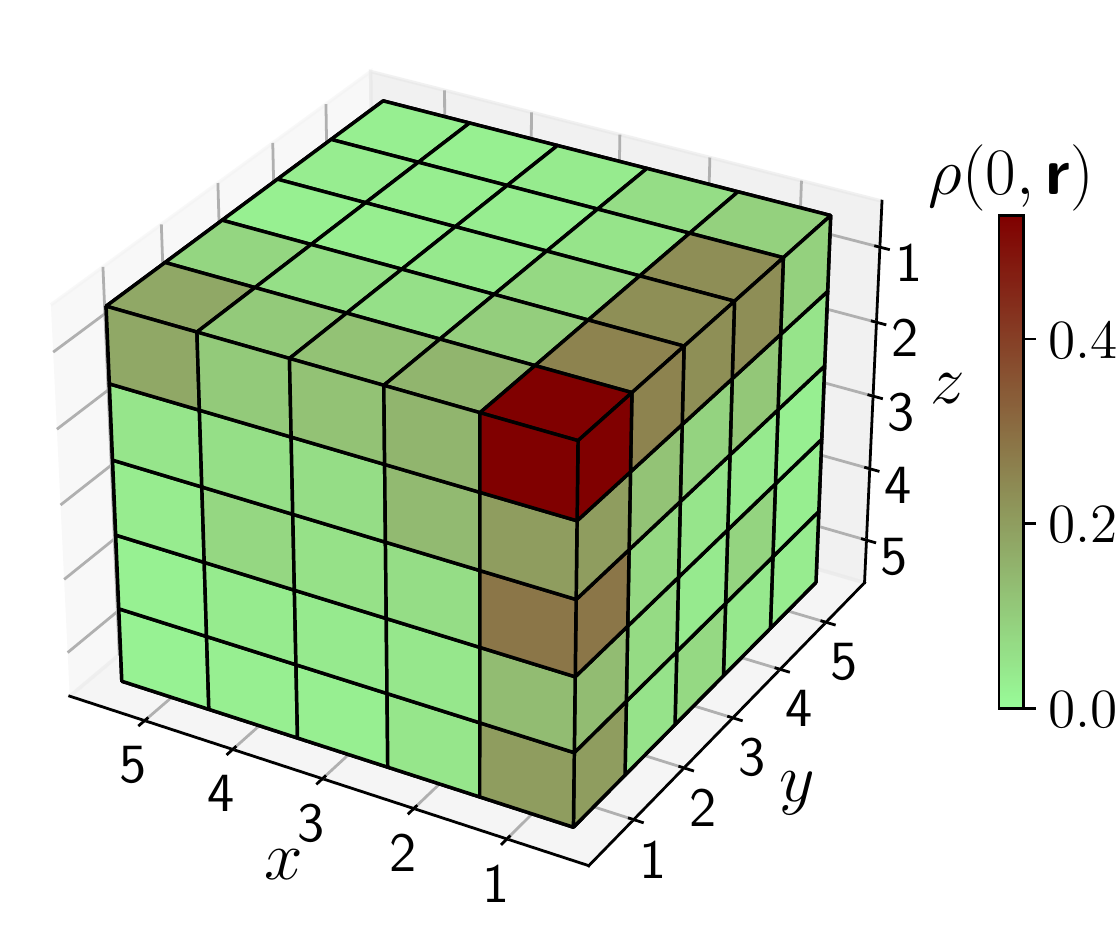}};
      \node[draw = none, fill = none] at (\width,0){\includegraphics[width=\width]{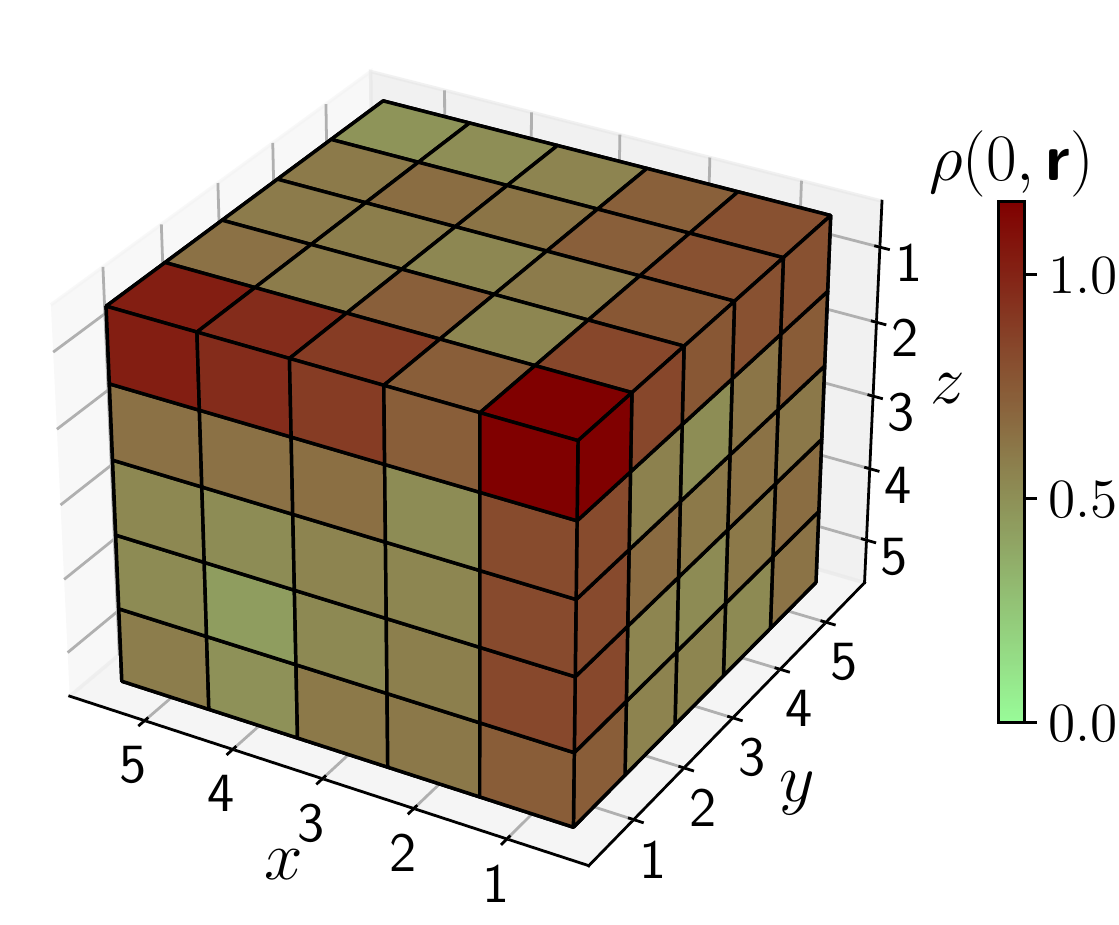}};

      \node at (-0.4*\width,1.5) {\textbf{(a)}};
      \node at (0.6*\width,1.5) {\textbf{(b)}};
    \end{tikzpicture}
  }
  \caption{
    Local density of states at zero-energy as function of unit-cell number $\textbf{r}$, in a corner of a system with size $L = 30$.
    The kernel polynomial method was used with a single random state trace approximation and $N = 2^{10}$ moments, averaged over 200 disordered samples with disorder weight (a) $W = 3$ (phase II) and (b) $W = 4$ (phase III).
  }
  \label{fig_ldos}
\end{figure}

\begin{figure*}
  \centering
  \resizebox{\textwidth}{!}{
    \def\width{2.3in}
    \begin{tikzpicture}
      \node[draw = none, fill = none] at (-\width,0){\includegraphics{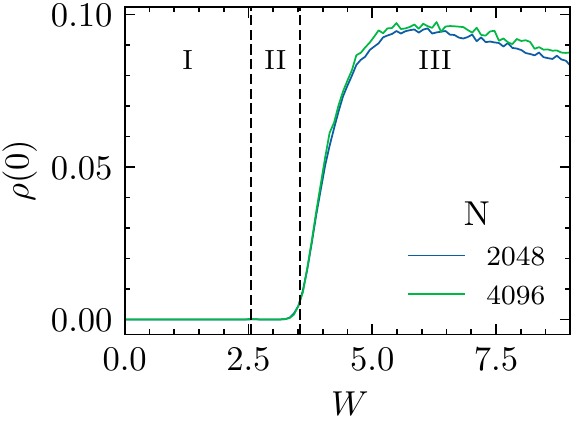}};
      \node[draw = none, fill = none] at (0,0){\includegraphics{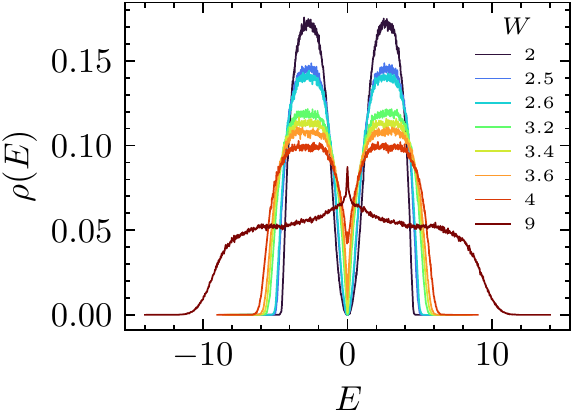}};
      \node[draw = none, fill = none] at (\width+6,0){\includegraphics{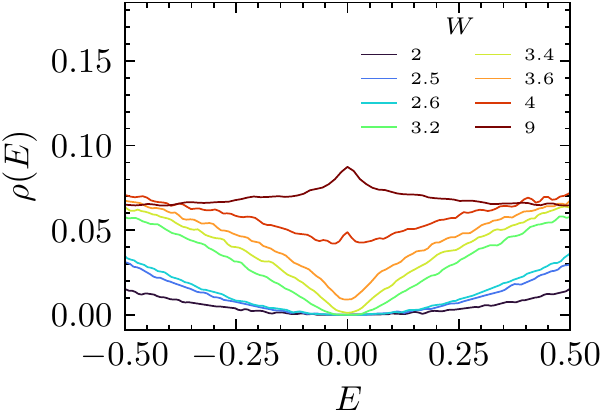}};
      \node at (-0.40\linewidth,1.8) {\textbf{(a)}};
      \node at (-0.07\linewidth,1.8) {\textbf{(b)}};
      \node at (0.26\linewidth,1.8) {\textbf{(c)}};
    \end{tikzpicture}
  }
  \caption{Density of states $\rho(E)$, computed with the kernel polynomial method for a system size $L = 80$.
  (a) $\rho(E=0)$ as function of disorder strength $W$. The two curves shown are for different choices of the number of Chebyshev moments $N$.
  (b) $\rho(E)$ computed with $N = 2^{13}$ moments, for selected values of $W$, with a zoomed-in view around the zero-energy region in (c).
  }
  \label{fig_dos}
\end{figure*}
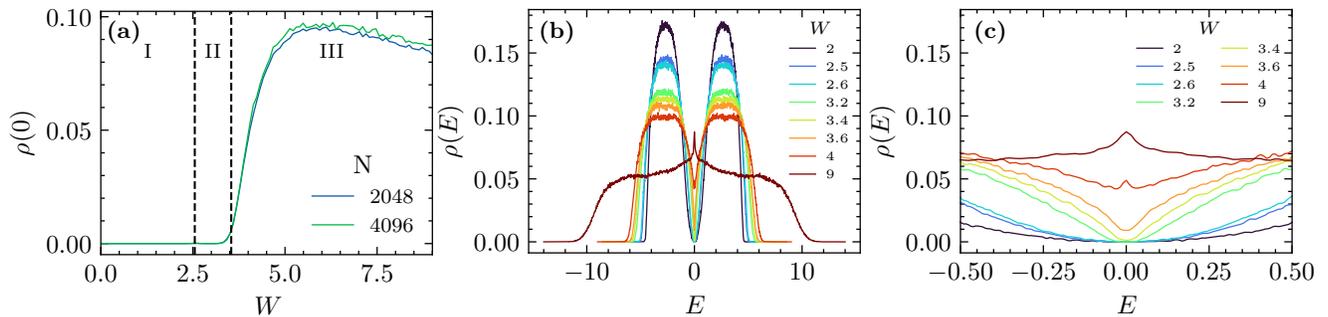

\begin{table}
  \centering
  \resizebox{\linewidth}{!}{
    \begin{tabular}{|c|c|c|c|c|}
      \hline
      Phase & I: GI & II: TOTAI & III: DM & IV: AI \\   
      \hline
      Topology & Trivial & Non-trivial & Trivial & Trivial\\
      \hline
      Spectrum & Gapped & Gapped & Gapless & Gapless \\
      \hline
      Zero-energy states & Localized & Localized & Extended & Localized \\
      \hline
      $W_c$ & $-$ & $2.55(20)$ & $3.54(3)$ & $24(2)$  \\
      \hline
    \end{tabular}
  }
  \caption{Summary of all the phases observed in the model for $\gamma = 1.1$: trivial gapped insulator (GI), third-order topological Anderson insulator (TOTAI), diffusive metal (DM) and Anderson insulator (AI); with the respective topological, spectral and localization properties.}
  \label{tab_phases}
\end{table}

Starting from a trivial insulator in the clean limit, $\gamma = 1.1$, we found four different phases as a function of disorder strength $W$, that are summarized in Fig.$\,$\ref{fig_phases_scheme} and Tab.$\,$\ref{tab_phases}.
In the next sections, we detail the properties of each phase.

\subsection{Topological phase diagram}\label{sec_res_top}
In Fig .$\,$\ref{fig_phase_diag}(a), the phase diagram for the topological invariant $Q$ is shown.
Due to the large finite-size effects, extrapolations to $L\rightarrow \infty$ were performed.
For the extrapolations, three linear fits were performed for $Q(L^{-1})$, for the 5 largest values of $L$ (\textcolor{plotgreen}{green}), for the 10 largest values of $L$ (\textcolor{plotorange}{orange}) and also for all values of $L$ (\textcolor{plotred}{red}), as shown in Fig.~\ref{fig_phase_diag}(b). 
The extrapolated value of $Q\left(L^{-1} \rightarrow 0\right)$ is the average result of the three fits.
Starting from a topologically trivial phase I, at the critical disorder $W_c^{\textnormal{II}} = 2.55(20)$ an abrupt increase in $Q$ occurs, indicating the start of phase II. 
$W_c^{\textnormal{II}}$ is precisely determined by the lowest $W$ for which $Q$ increases and a relatively large error is considered to take into account finite-size effects.
For this phase II, extrapolated values of $Q$ are compatible with $Q=1$ within error bar, signaling a topologically non-trivial phase. As shown in Fig.~\ref{fig_phase_diag}(c),  the gap closes and reopens at $W_c^{\textnormal{II}}$, further pointing to a transition into a tropological phase. Since this phase was induced by disorder, we dubbed it a TOTAI. However, it is important to note that the system is gapped in this phase, as is evident from Fig.~\ref{fig_phase_diag}(c).

Further increasing disorder, the system transitions to phase III, where extrapolated values of $Q$ are compatible with $Q=0$, indicating that it is topologically trivial. These results are also compatible with the zero-energy LDOS  shown in Fig.~\ref{fig_ldos}, revealing the existence of localized protected corner states in the TOTAI phase II, and their absence in the trivial phase III.
We estimated $W_c^\textnormal{III}$ from the extrapolation to $L\rightarrow \infty$ (analogously to Fig. \ref{fig_phase_diag}(b)), of the crossing between the energy gap and the mean level spacing (not shown) of the 50 states closer to $E=0$, resulting in $W_c^{\textnormal{III}} = 3.51(3)$. 
This estimation will be compared to a different one based on localization properties in section~\ref{sec_res_loc}.

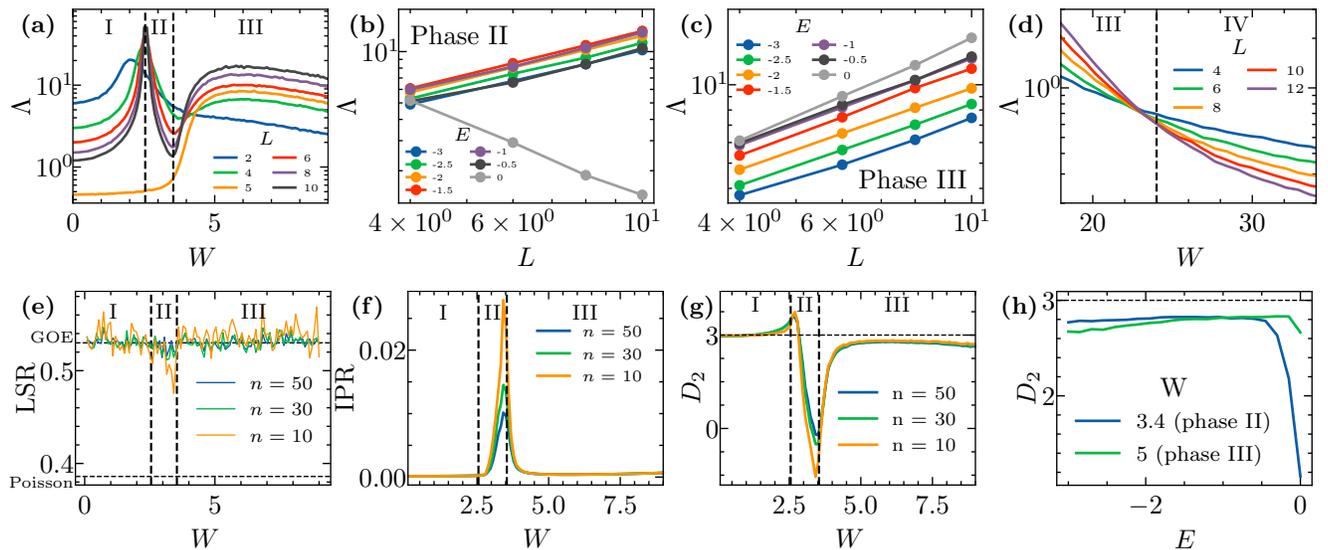
\begin{figure*}
  \centering
  \resizebox{\textwidth}{!}{
    \def\width{1.7in}
    \begin{tikzpicture}
      \node[draw = none, fill = none] at (0,0){\includegraphics{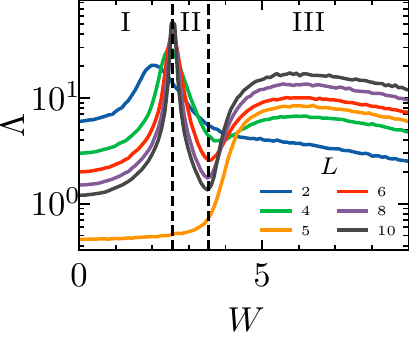}};
      \node[draw = none, fill = none] at (\width,0){\includegraphics{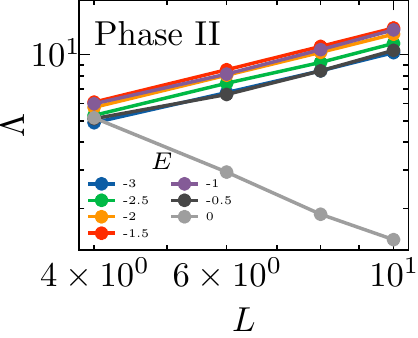}};
      \node[draw = none, fill = none] at (2*\width,0){\includegraphics{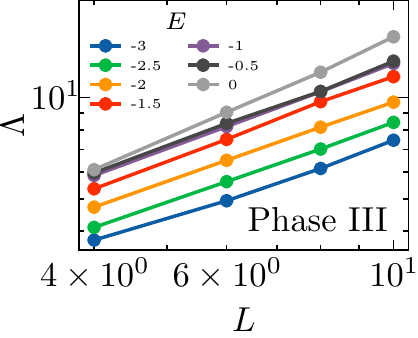}};
      \node[draw = none, fill = none] at (3*\width,0){\includegraphics{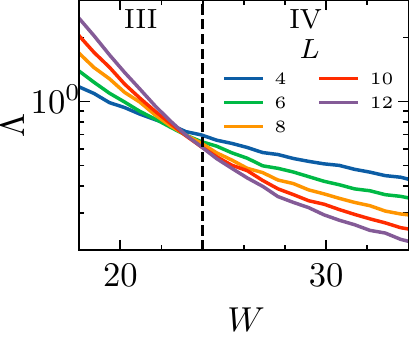}};

      \node[draw = none, fill = none] at (0,-3.7){\includegraphics{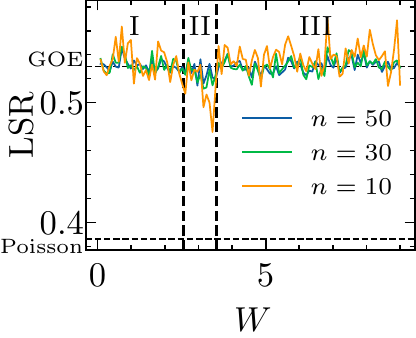}};
      \node[draw = none, fill = none] at (\width,-3.7){\includegraphics{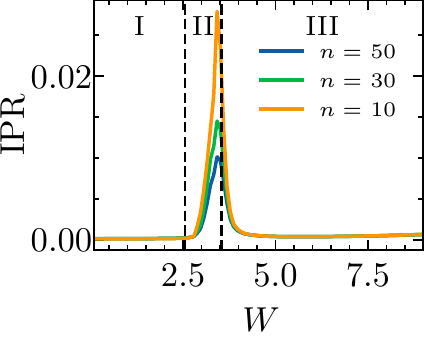}};
      \node[draw = none, fill = none] at (2*\width,-3.7){\includegraphics{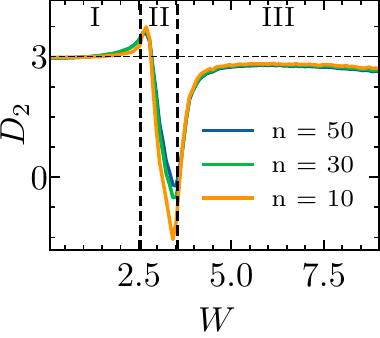}};
      \node[draw = none, fill = none] at (3.02*\width,-3.7){\includegraphics{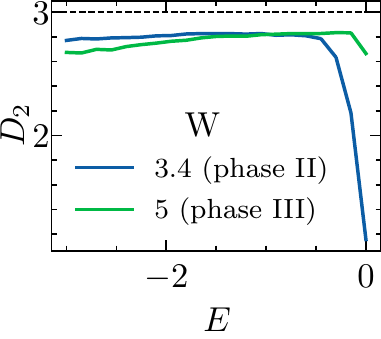}};

      \node at (-0.4*\width,1.5) {\textbf{(a)}};
      \node at (0.6*\width,1.5) {\textbf{(b)}};
      \node at (1.6*\width,1.5) {\textbf{(c)}};
      \node at (2.6*\width,1.5) {\textbf{(d)}};

      \node at (-0.4*\width,-2.2) {\textbf{(e)}};
      \node at (0.6*\width,-2.2) {\textbf{(f)}};
      \node at (1.6*\width,-2.2) {\textbf{(g)}};
      \node at (2.58*\width,-2.2) {\textbf{(h)}};
    \end{tikzpicture}
  }
  \caption{
    Normalized localization length $\Lambda$ from the transfer matrix method at (a,d) E = 0 versus $W$ for distinct $L$, and versus $L$ for distinct energies at (b) $W = 3.4$ and at (c) $W = 5$.
    (e) LSR, (f) IPR and (g) fractal dimension $D_2$ for $n$ eigenstates around zero-energy from exact diagonalization, as a function of $W$.
    (h) $D_2$ versus E for distinct $W$ for $n = 10$ eigenstates around $E$.
    Averages were taken over 200 realizations of disorder.
    In (e) and (f), $L =20$. 
    $D_2$ was computed by fitting using the sizes $L \in \{10,12,\dots,20\}$ in (f) and $L \in \{4,6,\dots,16\}$ in (g).
  }
  \label{fig_loc}
\end{figure*}

\subsection{Density of States}\label{sec_res_dos}

In Fig.$\,$\ref{fig_dos} we show the DOS for different disorder strengths.
We note that at the topological transition from phase I to II, although the gap closes, $\rho(0)$ is always zero [see Fig.~\ref{fig_dos}(a)], behaving as $\rho(E) \sim E^2$ around $E=0$ [see Fig.~\ref{fig_dos}(c)], as it would for a clean system with a Dirac cone (which is the case of the clean 3D-BBH model in the topological transition point). 
This was verified by performing a linear fit in a log-log plot of the curves in Fig.~\ref{fig_dos}(b) of positive $E$ values close to $E = 0$ for $W \in \{2.5, 2.6\}$, which rendered slopes compatible with two (not shown).
In phase III, the energy gap closes again and  $\rho(0)$ becomes finite. 
The DOS starts to become peaked around $E=0$ for large $W$. Whether this finite DOS at the Fermi level ($E=0$) is associated with a diffusive metal or an Anderson insulator is discussed next.

\subsection{Localization properties}\label{sec_res_loc}

In Fig. \ref{fig_loc}(a), we plot the normalized localization length $\Lambda$ along the $z$-direction at $E=0$. The calculations of $\Lambda$ along other directions yielded quantitatively identical results. 
We can see that $\Lambda$  decreases with $L$ in phases I and II. This is because the system is gapped and the wave function can therefore only propagate through (evanescent) localized modes at $E=0$. 
At the topological phase transition, however, $\Lambda$ becomes $L$-independent, as expected. 
In phase III, the system is gapless and has extended states at $E=0$ since $\Lambda$ increases with $L$, as expected for a diffusive metal. 
We also note that for energies where the DOS is finite,  the eigenstates are extended in phases I-III, as supported in Figs$\,$\ref{fig_loc}(b,c).

For large $W$, we see another phase transition at $W_c^{\textnormal{IV}} = 24(2)$ to a phase IV where $\Lambda$ again decreases with $L$, Fig. \ref{fig_loc}(d). In this case, even though the system is gapless, the bulk extended states become localized at $E=0$.
In fact, localization occurs at all energies and corresponds to the standard Anderson transition \cite{anderson1958absence,lagendijk2009fifty,evers2008anderson}.

In order to make an additional independent estimation of the critical point for the transition from phase II to III, we also analyzed the crossing points between curves of adjacent $L$ in Fig.$\,$\ref{fig_loc}(a).
Fitting the crossing points analogously to what was done in \ref{fig_phase_diag}(b), we extrapolate $W_c^{\textnormal{III}} = 3.56(3)$ in the thermodynamic limit.
This is compatible with the result obtained in section \ref{sec_res_top} and their average is presented in Tab. \ref{tab_phases}.

We now turn to the LSR analysis. In Fig.$\,$\ref{fig_loc}(e), we present the LSR for eigenenergies around $E = 0$, where we had to disregard some abnormally large outlier spacings created due to finite-size effects (they correspond to spacings between sets of degenerate states in the clean limit).
The LSR in phase III follows GOE statistics, completing the proof that phase III is a diffusive metal. 
In phases I and II, where we access the statistics of the gap edge, the states mostly follow the GOE ensemble for diffusive and extended states.
However, as we approach the transition point $W^{\mathrm{III}}_c$, there is a sudden decrease in the LSR, especially at lower $n$ (closer to the gap edge). To better understand this result, we calculated the IPR and the fractal dimension, which we discuss next.

For the gapped phases, we computed the average IPR for eigenstates at the gap edge, as for the LSR. 
In Fig.$\,$\ref{fig_loc}(f), we can see that the IPR is small in phases I and III, which, in combination with the obtained fractal dimension $D_2\approx 3$ in Fig.$\,$\ref{fig_loc}(g), indicates that the eigenstates closer to $E=0$ are extended. 
We also observe in Fig.$\,$\ref{fig_loc}(f) that the IPR becomes larger in phase II, peaking close to the transition II $\rightarrow$ III. 
This is concomitant with the fractal dimension results in Fig.$\,$\ref{fig_loc}(g), where it can be seen that $D_2 \approx 0$ close to the transition, suggesting the presence of localized gap-edge states right before the gap closes. This correlates with the sudden drop of the LSR. However, there are still some discrepancies between the results for the LSR and fractal dimension (the LSR is still significantly away from Poisson), that we attribute to strong finite-size effects in phase II.
Fig.$\,$\ref{fig_loc}(h) further shows that in phase III the states are extended for any energy, while in phase II the states are only localized close to $E = 0$, at the gap edges. 
These localized states are likely related with Lifshitz tails, whose exponentially suppressed DOS in the thermodinamic limit justifies the strong finite size effects, especially for the LSR results.

\subsection{Self-Consistent Born Approximation}

\begin{figure}
  \centering
  \includegraphics[width=3in]{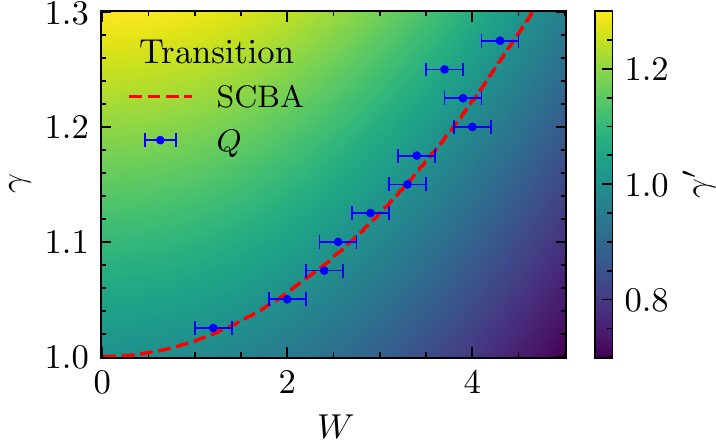}
  \caption{
    Effective renormalizaded intra-cell hopping amplitude $\gamma'$, computed through the SCBA, as a function of the clean hopping amplitude $\gamma$ and the disorder strength $W$.
    The topological transition curve at $\gamma' = 1$ is shown in red and the transition numerically extracted from the topological invariant $Q$ is shown in the blue points.
  }
  \label{fig_scba}
\end{figure}

Disorder is introduced into the system in the form of added intra-cell hopping amplitudes at each unit cell $\textbf{r}$, that is,
\begin{equation}
  V_\textbf{r} = \sum_{\alpha = 1}^{12} V_{\textbf{r}, \alpha} U_{\alpha} \, ,
\end{equation}
where $\alpha(i,j) \in \{1,\dots,12\}$ is a bijection between the index of an edge $\alpha$ and the indexes $i,j$ of the adjacent corners.
$V_{\textbf{r}, \alpha(i,j)} = W \Delta_r^{ij}$ are the hopping strengths and 
\begin{equation}
    \left[ U_{\alpha(i,j)}\right]_{mn} = \frac{1}{\gamma} \left[ \Gamma_0 \right]_{mn} (\delta_{mi} \delta_{nj} + \delta_{mj} \delta_{ni})
\end{equation}
are the matrix elements of each separate intra-cell hopping term.
Since the disorder is uncorrelated,
\begin{equation}
  \langle V_{\textbf{r}, \alpha} \rangle = 0 \ , \  \langle V_{\textbf{r}, \alpha} V_{\textbf{r}', \beta} \rangle = \frac{W^2}{12} \delta_{\textbf{r}\textbf{r}'} \delta_{\alpha\beta} \ .
\end{equation}

Under the Self-Consistent Born approximation (SCBA)\cite{bruus2004many, groth2009theory, yang2021higher, mavsek1986electronic}, the effective Bloch Hamiltonian at $E = 0$ is $H_{\textnormal{eff}}(\textbf{k}) = H_0(\textbf{k}) + \Sigma(E = 0)$, where the self-energy $\Sigma$ is computed self-consistently through the following equation,
\begin{equation}
  \Sigma(E) = \frac{W^2}{12 (2\pi)^3} \int_{BZ} d^3 \textbf{k} \sum_{\alpha = 1}^{12} U_\alpha G U_\alpha \, ,
\end{equation}
where $G = \left[(E+i0^+) \mathds{1} -H_0(\textbf{k}) - \Sigma(E)\right]^{-1}$ is the Green's function.
Numerically, we find that $\Sigma(0) = - \Gamma_0 \sigma / \gamma$, $\sigma \in \mathbb{R}$. 
In the effective Hamiltonian, this amounts to a normalization of the intra-cell hopping strengths $\gamma \rightarrow \gamma' = \gamma - \sigma$.
Since the effective model still corresponds to the 3D-BBH clean model, the topological (trivial) phase occurs for $\gamma' < 1 (>1)$. 
In Fig. \ref{fig_scba}, we observe that the topological transition curve predicted by the SCBA agrees very well with the one computed numerically from the topological invariant $Q$.

\section{Conclusions}\label{sec_conclusions}

In summary, we have discovered the first example of a third-order topological Anderson insulator, induced by chiral symmetry preserving disorder.
The TOTAI phase is characterized by a quantized quadrupole moment on the boundaries of the 3D system, that corresponds to a quantized bulk octupole moment, and by topologically protected localized corner states. Remarkably, the topological transition to the TOTAI phase is captured with great accuracy by the self-consistent Born approximation, up to very large disorder strengths.

Our findings can be tested experimentally in different metamaterials where disorder can be tuned, such as mechanical metamaterials \cite{serra2018observation}, electric circuits  \cite{imhof2018topolectrical,PhysRevLett.126.146802,PhysRevB.100.201406} or photonic waveguides \cite{Dutt2020}.

Finally, we note that in contrast to the disordered 2D BBH model \cite{yang2021higher}, we have not found a gapless HOTAI in 3D.
This raises an interesting open question for future research: do gapless TOTAIs exist?

\section{Acknowledgments}

This work has been partially funded by the ERC Starting Grant 101042293 (HEPIQ) (H.L.). 
The authors MG and PR acknowledge partial support from Fundação para a Ciência e Tecnologia (FCT-Portugal) through Grant No. UID/CTM/04540/2019.  EVC acknowledge partial support from FCT-Portugal through Grant No. UIDB/04650/2020. MG acknowledges further support from FCT-Portugal through the Grant SFRH/BD/145152/2019.
We finally acknowledge the Tianhe-2JK cluster at the Beijing Computational Science Research Center (CSRC), the Bob|Macc supercomputer through computational project project CPCA/A1/470243/2021 and the OBLIVION supercomputer, through projects HPCUE/A1/468700/2021, 2022.15834.CPCA.A1 and 2022.15910.CPCA.A1 (based at the High Performance Computing Center - University of Évora) funded by the ENGAGE SKA Research Infrastructure (reference POCI-01-0145-FEDER-022217 - COMPETE 2020 and the Foundation for Science and Technology, Portugal) and by the BigData@UE project (reference ALT20-03-0246-FEDER-000033 - FEDER and the Alentejo 2020 Regional Operational Program. Computer assistance was provided by CSRC's, Bob|Macc's and OBLIVION's support teams. 

\bibliographystyle{apsrev4-2}
\bibliography{biblio}

\end{document}